\documentclass[runningheads]{llncs}
\usepackage[T1]{fontenc}
\usepackage{graphicx}
\usepackage{subcaption}
\usepackage{amsmath}
\usepackage{amssymb}
\usepackage{float}
\usepackage{placeins}
\usepackage{algpseudocode}
\usepackage{algorithm}
\usepackage{cite}

\usepackage{bm}
\usepackage{multirow}
\usepackage{multicol}

\begin{document}
\title{Redefining POI Popularity: Integrating User Preferences and Recency for Enhanced Recommendations}
\titlerunning{Redefining POI Popularity: Integrating User Preferences and Recency}
\author{Alif Al Hasan\inst{1}\orcidID{0009-0000-3752-616X} \and
Md Musfique Anwar\inst{1}\orcidID{0000-0001-5159-1865} \and
Muhammad Arifur Rahman\inst{2}\orcidID{0000-0002-6774-0041}}
\authorrunning{A.A. Hasan et al.}
\institute{Department of Computer Science and Engineering, Jahangirnagar University, Savar, Dhaka-1342, Bangladesh\\
\email{\{alif.stu2017, manwar\}@juniv.edu} \and
Department of Computer Science, Nottingham Trent University, NG11 8NS, UK\\
\email{arif.rahman@ntu.ac.uk}}
\maketitle
\begin{abstract}
The task of point-of-interest (POI) recommendation is to predict users' immediate future movements based on their previous records and present circumstances. Popularity is considered as one of the primary deciding factors for selecting the next place to visit. Existing approaches mainly focused on the number of check-ins to model the popularity of a POI. However, not enough attention is paid to the temporal impact or number of people check-ins for a particular POI. Thus, to prioritize more on recent check-ins, we propose \textit{recency}-oriented definition of POI's popularity by considering the \textit{temporal} effect of the POIs, the number of check-ins, as well as the number of users who registered in those check-ins. Our experimental results on real dataset show the efficacy of the proposed approach.

\end{abstract}
\section{Introduction}
Over the past few years, location-based social networks (LBSNs) have experienced significant developments, including Yelp, Foursquare, and others. By checking-in at places of interests, users may share their locations and experiences with their friends. The visited point of interest (POI) together with any accompanying contexts (e.g., categories, timestamp, GPS) that define the user's movement are typically included in a check-in record. The enormous amount of check-in data produced by millions of users in LBSNs offers a fantastic chance to investigate the underlying patterns in user check-in behavior. Numerous point-of-interest and travel recommendation systems take advantage of this vast quantity of information.

When choosing the next point of interest (POI), nearly all trip and POI recommender systems employ popularity as one of their measures. The majority of them characterize the number of check-ins as a POI's popularity indicator. However, determining popularity only by check-in numbers isn't necessarily the best course of action. This provides us with a few things to improve.

First, a point's level of popularity varies with time. Points A and B, for instance, have nearly equal numbers of check-ins. While the majority of the check-ins in point-B were done a few years ago, the majority of the check-ins in point-A were made recently. Point-B was apparently popular back then, based on this scenario, but choosing point-A over point-B makes a lot more sense now. 

Second, a place's popularity among a certain demographic does not guarantee that it will be favored by everyone. For instance, a significant number of individuals go to point-A regularly, let's say on a weekly basis, whereas only a handful of people meet at place-B but they go there every day. Hence, even though there are exactly the same number of check-ins at these two locations, it is very possible that a newcomer would select point-A over point-B because point-A is preferred by a wider range of individuals. 

Our next POI selection problem will include redefining popularity in light of these observations. Our suggested strategy essentially incorporates the new definition of popularity into the GETNext method \cite{Yang} which builds a graph enhanced transformer model by combining the user's general preference, spatio-temporal context, global transition pattern, and time-aware category embedding to predict the user's next move. To extract the relevant information from the raw data, we first preprocessed the data. Next, using the user's current trajectory as a guide, we apply the modified version of the GETNext algorithm to the processed data to determine the next top-\textit{k} POI recommendation. In short, the following sums up our contributions:
    \begin{itemize}
        \item We proposed a new definition of popularity for the next POI or trip recommendation system.
        \item We demonstrated how the temporal aspect affects the popularity of POIs over time and why it is advisable to choose the POI that is more widely accepted.
        \item We performed experiments on a real dataset to demonstrate the efficacy of our proposed method.
    \end{itemize}

In Section 2, we reviewed several studies that are relevant to our work. We made an effort to summarize their approach and set it apart from ours. In-depth analysis of the subtleties of problem formulation is provided in Section 3. All of the details of our suggested methodology and initial concepts are covered in Section 4. Discussion about our experimental setup, dataset, and findings are revealed in Section 5. A summary of our overall contributions are included in Section 6 along with wrapping up the work and outlining our next research goals.

\section{Related Work}
There are several methods available for POI or trip recommendation systems, including matrix factorization, recurrent neural networks (RNN), and graph-based approaches. Among the tactics employed in early research are Markov chains, which have been frequently used in other sequential recommendation problems. For example, Cheng et al. \cite{Cheng} proposed next points of interest in their groundbreaking study by combining FPMC (Personalized Markov Chains) \cite{Rendle} with a matrix factorization technique. In a similar vein, Zhang et al. \cite{Zhang} proposed to use an additive Markov chain to simulate the sequential transitive influence. When it comes to modeling sequence data, these early techniques are not as effective as deep neural network models.

Deep learning and sophisticated embedding techniques have been the subject of recent study \cite{Feng}. RNN variations were proposed \cite{Liu, Sun, Wu0, Wu, Zhao, Zhao2} to reflect the temporal dynamics and sequential correlations. Spatial temporal recurrent neural networks (ST-RNN) were introduced by Liu et al. (2016) \cite{Liu}. These networks integrated temporal and spatial contexts into RNN layers. Specifically, time transition matrices describe temporal context, while geographic distance transition matrices characterize spatial contexts. The participants' short- and long-term preferences were also approximated using LSTM. For short-term trajectory mining, the authors of PLSPL \cite{Wu} and LSPL \cite{Wu0} trained a general embedding layer to get user preferences and a conventional LSTM model. DeepMove \cite{deepmove} proposed an attention model to capture the multi-level periodicity pattern and a recurrent neural network to simulate the sequential transitions.

Graph-based approaches, such as those utilizing location-based social networks (LBSN), offer a potent paradigm, particularly for traditional (non-sequential) recommendation tasks. Yuan et al. \cite{Yuan} created the geographical-temporal influences aware graph (GTAG), a tripartite graph including POI, session, and user nodes. The authors of a recent work \cite{ijcai2021p206} randomly picked previous and next check-ins from different check-in sequences for each POI in the provided check-in sequence in order to train an embedding of the current POI. Since the intention was to capture local (one-hop) transition of POIs, the embedding does not particularly reflect the global (multi-hop) transition patterns. In their subsequent POI recommendation, Yang et al. \cite{Yang} used graph-based approaches. To recommend the next point of interest, the authors deliberately employed a uniform graph structure in their methodology. All of the POIs' global patterns were represented by a single graph structure, or modeled trajectory flow map. The authors used graph-based learning to encode generic transitional information about POIs in the next POI recommendation problem.

We find that the quantity of users checking in and the \textit{recentness} of check-in records have a significant impact on a POI's popularity. Most current study approaches considered only the number of check-ins and paid little attention to the number of users checking in. We propose a novel definition of popularity for a POI that considers both the total number of check-ins and the number of people who do not check-in. Furthermore, it prioritizes the current trend in popularity over those that were popular a while ago but aren't as popular right now.

\section{Problem Formulation}
Now, let's explore the key ideas presented in the paper. Starting with the concept of users, represented by \(U\), we create a collection \(U = \{u_1, u_2, \ldots, u_M\}\), in which \(M\) is the total number of users. Points of Interest (POIs) are the next category and include various locations such as restaurants, hotels, coffee shops, parks, malls, clothing stores, bus stations, airports, etc. They are represented by the symbol \(P = \{p_1, p_2, \ldots, p_N\}\), where \(N\) denotes the number of unique POIs. The timestamps, representing discrete points in time, are stored in the set \(T = \{t_1, t_2, \ldots, t_K\}\), where \(K\) represents the total number of timestamps.

Within set \(P\), every POI \(p\) is specified by a tuple \(p = \langle \text{cat}, \text{freq}, \text{lat}, \text{lon} \rangle\), containing its category (cat), frequency of visits (freq), and geographical coordinates (latitude (lat) and longitude (lon)). The category, denoted by \(\text{cat}\), correlates to preset categories such as "train station" or "restaurant."

\textbf{Definition 3.1 (Check-in)}: A check-in, represented by a tuple \(q = \langle u, p, t \rangle \in U \times P \times T\), signifies that a user \(u\) visited POI \(p\) at timestamp \(t\).

The sequence of a user's check-in activities, \(Q_u = (q_u^1, q_u^2, q_u^3, \ldots)\), is an ordered collection of these events, where each \(q_u^i\) represents the \(i\)-th check-in record.

\textbf{Definition 3.2 (Check-in Set)}: The check-in sequences for each user are included in the collection/set denoted as \(Q_U = \{Q_{u_1}, Q_{u_2}, \ldots, Q_{u_M}\}\).

During data preprocessing, we divide each user's check-in sequence \(Q_u\) into successive trajectories, represented as \(Q_u = S_u^1 \oplus S_u^2 \oplus \ldots\), concatenating them with \(\oplus\). These trajectories, which differ in length, represent a sequence of check-ins that occur over predetermined periods, such as an entire day.

The main objective of our work is to predict a user's next trips to POIs by using their past check-in history and present trajectory. For a given user \(u_i \in U\), given a set of historical trajectories \(\{S_u^i\}_{i \in \mathbb{N}, u \in U}\) and a current trajectory \(S' = (q_1, q_2, \ldots, q_m)\), our goal is to determine the most likely future POIs (\(q_{m+1}, q_{m+2}, \ldots, q_{m+k}\)) to be visited by \(u_i\), where \(k \geq 1\) is typically set to 1.

%
%
%

%\section{Our Popularity Definition}
\iffalse
In our case, popularity is represented by a multivariate formula that attempts to capture different aspects of user interaction. Now let's explore its specifics:

\begin{equation}
    \label{eq:pop}
    \begin{split}
        \text{Popularity} = & \beta \left( \alpha \cdot C_{\text{user}}^{\text{recent}} + (1 - \alpha) \cdot C_{\text{check-in}}^{\text{recent}} \right) \\
        & + (1 - \beta) \left( \alpha \cdot C_{\text{user}}^{\text{past}} + (1 - \alpha) \cdot C_{\text{check-in}}^{\text{past}} \right)
    \end{split}
\end{equation}

The formula represents the number of recent user checkins as \(C_{\text{user}}^{\text{recent}}\) and the total number of unique users who checked in recently as \(C_{\text{check-in}}^{\text{recent}}\). On the other hand, the numbers of unique users who checked in and the total number of check-ins prior to the most recent ones are represented by \(C_{\text{user}}^{\text{past}}\) and \(C_{\text{check-in}}^{\text{past}}\), respectively. The weighting factors defined by the parameters \(\alpha\) and \(\beta\) indicate the relative importance of user count compared to check-in count and current records compared to older ones. \(\alpha\) and \(\beta\) are both restricted to the interval \textbf{\(0 \leq \alpha, \beta \leq 1\)}, which guarantees that they follow real-number constraints.
\fi

%
%
%

\section{Revised GETNext Method for the Next POI Recommendation}

We modified the GETNext model \cite{Yang} by introducing a new measurement for POI popularity. In our case, popularity is represented by a multivariate formula that attempts to capture different aspects of user interaction. Now let's explore its specifics:

\begin{equation}
    \label{eq:pop}
    \begin{split}
        \text{Popularity = } & \beta \left( \alpha \cdot C_{\text{user}}^{\text{recent}} + (1 - \alpha) \cdot C_{\text{check-in}}^{\text{recent}} \right) \\
        & + (1 - \beta) \left( \alpha \cdot C_{\text{user}}^{\text{past}} + (1 - \alpha) \cdot C_{\text{check-in}}^{\text{past}} \right)
    \end{split}
\end{equation}

The formula represents the number of recent user checkins as \(C_{\text{user}}^{\text{recent}}\) and the total number of unique users who checked in recently as \(C_{\text{check-in}}^{\text{recent}}\). On the other hand, the numbers of unique users who checked in and the total number of check-ins prior to the most recent ones are represented by \(C_{\text{user}}^{\text{past}}\) and \(C_{\text{check-in}}^{\text{past}}\), respectively. The weighting factors defined by the parameters \(\alpha\) and \(\beta\) indicate the relative importance of user count compared to check-in count and current records compared to older ones, respectively. The parameters \(\alpha\) and \(\beta\) are both restricted to the interval \textbf{\(0 \leq \alpha, \beta \leq 1\)}, which guarantees that they follow real-number constraints.

\begin{figure*}[t]
    \centering
    \includegraphics[width=\textwidth]{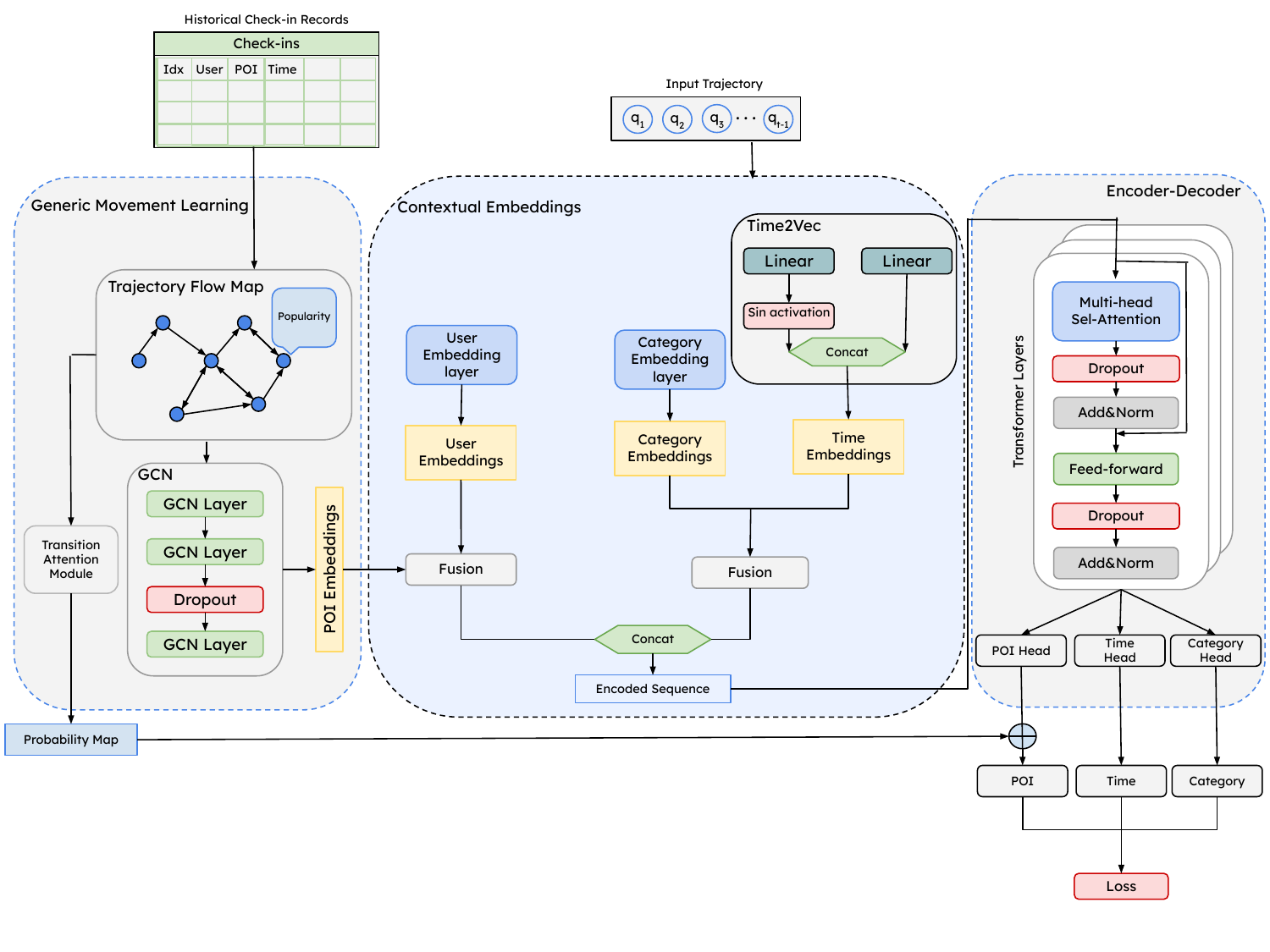}
    \caption{An overview of the GETNext model. \cite{Yang}} \label{fig1}
\end{figure*}

As seen in Figure ~\ref{fig1}, the GETNext model represents a comprehensive architecture made up of multiple essential components. The model primarily combines past trajectories, which is an important component described in the trajectory flow map (see Sec.~\ref{subsec:5-1}). The trajectory flow map has a major impact on recommendations in two important ways:

\begin{enumerate}
    \item The first step in the procedure is to use a trajectory flow map, which is an essential part of training a graph neural network (GNN). The GNN's goal is to generate Point of Interest (POI) embeddings—condensed representations that capture key properties like category, geographic location, and check-in frequency—through iterative training. These embeddings are essential for encoding users' generalized patterns of movement among different points of interest (POIs) in a particular region. The GNN gains the capacity to recognize important geographical and temporal correlations between various points of interest (POIs) by examining the trajectory flow map, which improves its comprehension and prediction of user preferences and behavior.
    \item An attention module is used in conjunction with the GNN training procedure to enhance the comprehension of users' movement patterns. A transition attention map, produced by this attention process, is a powerful tool for precisely modeling transition probabilities between various points of interest. The attention module is able to capture the dynamic nature of user movements and interactions with different points of interest over time by utilizing the adjacency matrix that is obtained from the trajectory flow map and incorporating node attributes as input. The generated transition attention map can be used to better predict and foster future POI visits by offering insightful information about the probability that users will move from one point of interest to another. This complete approach provides a framework for comprehending and analyzing user behavior in location-based services by merging attention-based transition modeling with GNN-based POI embeddings.
\end{enumerate}

Moreover, a variety of contextual modules are critical to our framework's ability to learn and interpret necessary encodings, which enhances its overall predictive power. These modules cover a range of features, including temporal encodings, POI category embeddings, and user embeddings; each has a unique function that is coupled to the others and is explained in more detail in Section~\ref{subsec:5-2}. Here is a quick explanation of how embedding modules function in our model:

\begin{itemize}
    \item We set out to jointly integrate POI embeddings from relevant trajectories with user embeddings in order to facilitate tailored recommendations. The adoption of a comprehensive strategy guarantees a sophisticated comprehension of user inclinations and enables the delivery of customized recommendations that align with the unique behaviors and preferences of each user.
    \item Furthermore, acknowledging the inherent significance of temporal dynamics in reshaping user preferences and actions, we explore the complex connection between POI category embeddings and time encodings. We want to comprehend the temporal dynamics that underlie users' preferences across various POI categories by carefully combining these components. To improve our recommendation system's accuracy in predicting outcomes, for example, identifying trends like often visiting train stations during rush hour requires a thorough integration of temporal signals with categorical embeddings.
\end{itemize}

A distinct check-in embedding vector is generated by merging user data, categories of points of interest (POIs), timestamps, and POI specifics in each check-in. As a result, a combination of these check-in embeddings represents each trajectory. After that, multilayer perceptron (MLP) heads and a transformer encoder are used to construct a prediction of a point of interest. The predicted POI is then adjusted using the obtained transition attention map in conjunction with a residual connection.

\subsection{Learning with Trajectory Flow Map}\label{subsec:5-1}
    \subsubsection{\textbf{Understanding Trajectory Flow Map}}
    An advanced method entails exploring the details of an attributed weighted directed graph, $G = (V, E, l, w)$. This graph functions as a trajectory flow map and is derived from an extensive set of historical trajectories $S = \{S_u^i\}_{i \in \mathbb{N}, u \in U}$, where:
    \begin{itemize}
        \item Points of Interest (POIs) correspond to the aggregation of nodes $V$.
        \item For every POI $p \in P$, there are features contained in $l(p)$. These include coordinates $(\text{lat, lon})$, category represented by $\text{category}$, and the frequency $\text{eq}$ of occurrence in trajectories in $S$.
        \item The edges that connect $p_1$ and $p_2$ in trajectory $S_u^i$ represented by $(p_1, p_2)$ are POI visits in succession.
        \item Each edge $(p_1, p_2)$ has a weight $w(p_1, p_2)$, which represents how often it occurs in any trajectory in $S$.
    \end{itemize}
    
    \subsubsection{\textbf{Learning POI Embedding}}
    Vectorized representations of Points of Interest (POIs) are obtained by using Graph Convolution Networks (GCN) with spectral convolution. Through the use of the trajectory flow map, this novel method encodes transition patterns and attributes, allowing for a more thorough comprehension of the overall dynamics of movement at each point of interest.
    
    \subsubsection{\textbf{Deciphering Transition Attention Map}}
    Use of a transition attention map, $\Phi$, allows for a detailed modeling of transition probabilities between POIs. Based on the intrinsic characteristics of the input node, this sophisticated map—which is calculated in the next transformer module—dynamically modifies the conclusions. The recommendation's results are greatly influenced by the $\Phi$ map, which illustrates the probability of moving from one POI to another.

\subsection{Contextual Embedding Module}\label{subsec:5-2}
Personalized next point-of-interest (POI) recommendations are built on top of the Contextual Embedding Module, which combines spatiotemporal contexts with user preferences. Two essential fusion components are included in this module:

\subsubsection{\textbf{POI-User Embeddings Fusion}}
In order to capture both the general patterns found in POIs and the user-specific behaviors encountered in previous check-in sequences, it is essential to combine user embeddings with POIs. Using a function $\text{fembed}(u)$, we first retrieve the user embedding $e_u$ in this case, which is represented as:
\begin{equation}
    e_u = \text{fembed}(u) \in \mathbb{R}^\Omega. \label{eq:user_embedding}
\end{equation}
User preferences and actions that are subtle are captured in this embedding.

Concatenating the POI embedding $e_p$ and the user embedding $e_u$ yields the fused embedding $e_{p,u}$, which is then used as follows:
\begin{equation}
    e_{p,u} = \sigma(w_{p,u} [e_p ; e_u] + b_{p,u}) \in \mathbb{R}^{\Omega \times 2}, \label{eq:fused_embedding}
\end{equation}
The activation function is indicated by $\sigma$, while the weights and bias are denoted by $w_{p,u}$ and $b_{p,u}$, respectively. To improve the model's capacity to collect customized recommendations, the concatenated vector $[e_p ; e_u]$ combines user attributes with points of interest.

\subsubsection{\textbf{Time-Category Embeddings Fusion}}
Categorical embeddings of POIs and temporal information recorded by Time2vector are combined in this fusion process. Taking into account the temporal aspect of user behavior, Time2vector efficiently encodes time values. In addition, an embedding layer is used for POI categories concurrently.

The following is the formulation of the fusion equation for time-category embeddings, $e_{c,t}$:
\begin{equation}
    e_{c,t} = \sigma(w_{c,t} [e_t ; e_c] + b_{c,t}) \in \mathbb{R}^{\Psi \times 2}, \label{eq:time_category_fusion}
\end{equation}
where the learnable weight vector is indicated by $w_{c,t}$, and the bias is represented by $b_{c,t}$. It is easier to integrate temporal and categorical data when $e_t$ and $e_c$ are concatenated.

The final embedding, $e_q$, is the combination of the POI-user and time-category embeddings, and it captures the core of a check-in operation. Each trajectory input consists of a series of check-in embeddings and is represented as $q = \langle p, u, t \rangle$, where POI $p$ is in category $c$. To facilitate precise POI recommendations, the transformer encoder further processes these embeddings to extract complex patterns and insights.
    
\subsection{Transformer Encoder and MLP Decoders}\label{subsec:5-3}
    \subsubsection{\textbf{Transformer Encoder}}
    A key element of our architecture is the transformer encoder, which consists of stacked layers with positional encoding and is essential to the next Point of Interest (POI) recommendation procedure. An input tensor \(X^{[0]} \in \mathbb{R}^{k \times d}\) is formed by concatenating historical check-in embeddings for each trajectory \(S_u\), where \(d\) is the embedding dimension. Normalization and residual connections are used in conjunction with fully linked networks and multi-head self-attention processes within each layer. The encoder layer produces an output designated as \(X^{[l+1]} \in \mathbb{R}^{k \times d}\) after a series of transformations.
    
    \begin{equation}
        S = X^{[l]}W_q (X^{[l]}W_k)^T \in \mathbb{R}^{k \times k}
    \end{equation}
    Using the query and key matrices \(W_q\) and \(W_k\) obtained from the encoder output, this equation calculates the similarity matrix \(S\).
    
    \begin{equation}
        S'_{i,j} = \frac{\exp(S_{i,j})}{\sum_{j=1}^{k} \exp(S_{i,j})}
    \end{equation}
    The attention weights obtained by softmax normalization of the similarity matrix \(S\) are represented by \(S'_{i,j}\) in this particular case.
    
    \begin{equation}
        \text{head}_1 = S'X^{[l]}W_v \in \mathbb{R}^{k \times \frac{d}{h}}
    \end{equation}
    The output of the first attention head is calculated using this equation.
    
    \begin{equation}
        Multihead(X^{[l]}) = [head_1; \ldots; head_h] \times W_o \in \mathbb{R}^{k \times d}
    \end{equation}
    The outputs from each attention head are combined into a single matrix by the multi-head attention mechanism.
    
    \begin{equation}
        X^{[l]}_{\text{attn}} = \text{LayerNorm}(X^{[l]} + \text{Multihead}(X^{[l]}))
    \end{equation}
    The application of layer normalization to the multi-head attention output and the sum of the input tensor is represented by this equation.
    
    \begin{equation}
        X^{[l]}_{\text{FC}} = \text{ReLU}(W_1X^{[l]}_{\text{attn}} + b_1)W_2 + b_2 \in \mathbb{R}^{k \times d}
    \end{equation}
    The layer normalization output is subjected to the feed-forward neural network modification.
    
    \begin{equation}
        X^{[l+1]} = \text{LayerNorm}(X^{[l]}_{\text{attn}} + X^{[l]}_{\text{FC}}) \in \mathbb{R}^{k \times d}
    \end{equation}
    By applying layer normalization to the total of the output from the feed-forward network and the attention mechanism, the transformer encoder layer's final output is obtained.
    
    \subsubsection{\textbf{MLP Decoders}}
    A key component in predicting the next POI, visiting time, and POI category is the multi-layer perceptron (MLP) decoders, which take input from the transformer encoder output. There are three different MLP heads that carry out these predictions. Combining the output from the POI head with the transition attention map yields the final recommendation. In other words, the time head models the intervals between check-ins, while the category head controls the forecasts for the following POI.
    
    \begin{equation}
        \hat{Y}_{\text{poi}} = X^{[l^*]}W_{\text{poi}} + b_{\text{poi}}
    \end{equation}
    Using the encoder output as a basis, this formula determines the anticipated next POI.
    
    \begin{equation}
        \hat{Y}_{\text{time}} = X^{[l^*]}W_{\text{time}} + b_{\text{time}}
    \end{equation}
    The encoder output is used in this calculation to determine the estimated visit time.
    
    \begin{equation}
        \hat{Y}_{\text{cat}} = X^{[l^*]}W_{\text{cat}} + b_{\text{cat}}
    \end{equation}
    Based on the encoder output, this equation indicates the next POI category that is predicted.
    
    \subsubsection{\textbf{Loss}}
    The most important component in the model's training process is the loss function, which takes into account several aspects of prediction accuracy. It combines different variables to offer an all-encompassing assessment, guaranteeing strong performance on a range of prediction tasks. In particular, the loss function includes cross entropy for both the temporal prediction and the point of interest (POI) category predictions, as well as mean squared error (MSE) for temporal prediction. Taken together, these metrics assess how well the model represents temporal dynamics and classifies points of interest and the categories that go along with them.

    A deliberate amplification method is used to handle the complexities of temporal prediction and preserve balanced gradients throughout optimization. Notably, in the final loss computation, the contribution of the temporal loss component is multiplied by a factor of 10. This purposeful modification seeks to ensure that temporal dynamics receive the proper attention in the training regimen and to lessen the possible dominance of other loss components.
    
    The final loss function can be mathematically represented by the following equation:
    
    \begin{equation}
        L_{\text{final}} = L_{\text{poi}} + 10 \times L_{\text{time}} + L_{\text{cat}}
    \end{equation}
    
    In this case, the individual loss contributions resulting from POI prediction, temporal prediction, and POI category prediction are denoted by the variables $L_{\text{poi}}$, $L_{\text{time}}$, and $L_{\text{cat}}$. The loss function offers a comprehensive assessment of model performance by combining these elements into a single framework and directing the optimization process in the direction of improved prediction accuracy and generalization capacity.

\section{Experiment}
\subsection{Dataset}
As part of our experiment, we carefully examined the FourSquare-NYC public dataset, which is an extensive collection of check-ins in New York City from April 2012 to February 2013 that was carefully collected and described by Dingqi et al. \cite{Dingqi}. Every entry in this list represents the user's name, the point of interest (POI) that was visited, the particular category that the POI falls under, the exact GPS coordinates indicating its location, and the timestamp indicating the interaction.

We started the process of data refining through eliminating users who had very few check-ins in the past, leaving only those who had an important record of at least ten actions that were recorded. In a similar manner, we removed POIs with less than 10 check-in records from the dataset to make sure our analysis was supported by reliable and significant data points.

After a thorough curation process, we changed our focus to segmenting users' check-in behaviors into coherent paths. A detailed picture of users' engagement patterns across time was provided by the division of these trajectories, which were distinguished by their temporal continuity and spatial diversity, into discrete segments separated by 24 hours. We acknowledged the natural fluctuations in check-in rates, therefore in order to maintain statistical integrity, we carefully examined trajectories where there was only one check-in. These were identified as outliers and removed from the dataset.

After carefully priming and preparing our dataset, we divided it into several subgroups for testing, validation, and training. The trajectory flow map—a fundamental artifact of our analytical journey—was carefully developed from the first 80\% of check-ins designated for training, the next 10\% for validation, and the last 10\% sacrosanct for testing.

One of the most important components of our technique was the strict exclusion criteria we used for evaluation, which made sure that any people or items of interest that we had not experienced in the training phase were not included in the performance assessment. As a result, our predictive models were protected from bias and overfitting, and a more sophisticated comprehension of their actual ability to predict was developed.

Significant statistics from the dataset are shown in Table~\ref{tab: dataset}.

\begin{table}[h]
    \caption{Dataset Statistics}\label{tab: dataset}
    \centering
    \begin{tabular}{|l|l|l|l|l|}
        \hline
        user & poi & cat & checkin & trajectory \\
        \hline
        1,075 & 5,099 & 318 & 104,074 & 14,160\\
        \hline
    \end{tabular}
\end{table}

\subsection{Evaluation Metrics}
We use advanced metrics to assess the performance of our recommendation system during the evaluation process. In particular, we explore two well-known metrics in the field of recommender systems: Mean Reciprocal Rank (MRR) and Accuracy@k (\text{Acc@k}). These metrics give important information about how well the model is performing, including how well it can recommend points of interest (POIs) to users.

To determine whether the real POI is among the top-\textit{k} recommended POIs, the Accuracy@k measure is used as an indicator of accuracy. It basically assesses how well the algorithm identified pertinent POIs from a given set of recommendations. This indicator is especially helpful for assessing how accurately and efficiently the system provides users with ideas that are relevant to their needs. Formally, the Accuracy@k is determined in this way:

\[
\text{Acc@k} = \frac{1}{m} \sum_{i=1}^{m} 1(\text{rank} \leq k)
\]

In this case, the total number of samples or trajectories in our dataset is indicated by \( m \). To indicate whether the rank of the real POI in the sorted list of recommendations is in the top k spots, the ranking indicator function \( 1 \) is utilized. The function produces \( 1 \) if the condition is met; \( 0 \) is produced otherwise.

The rating of the right recommendation inside the sorted list is taken into account by the Mean Reciprocal Rank (MRR), which provides a more comprehensive viewpoint. In contrast to Accuracy@k, which handles the top-k recommendations as an unordered list, MRR considers the exact location of the right recommendation within the ordered list. This statistic plays a crucial role in evaluating how well the system ranks and prioritizes pertinent points of interest. This is how the MRR is determined:

\[
\text{MRR} = \frac{1}{m} \sum_{i=1}^{m} \frac{1}{\text{rank}}
\]

The real next point of interest in the sorted list is indicated by the symbol \( \text{rank} \) in this case. A thorough assessment of the system's ranking performance is offered by the MRR, which is calculated by averaging the reciprocals of the ranks over all samples.

Better performance is essentially indicated by higher values of Accuracy@k and MRR, which show how well the system can appropriately recommend pertinent POIs to users. We can refine and optimize our recommendation system to increase user satisfaction and engagement by using these metrics, which are vital tools for evaluating the efficacy and usefulness of the system.

\subsection{Results}
We have conducted a number of experiments that have contributed significantly to our understanding of the complex dynamics of our model. Despite the trials, the fundamental pillars of our study, denoted by the variables $\alpha$ and $\beta$, as stated in equation \ref{eq:pop}, held firm. Our technique stayed largely unchanged across a resolute span of 20 epochs, which allowed our comparison study to remain coherent.

Check-ins sought solace in two discrete temporal domains—the recent and the past—amid the complex pattern of temporal dynamics. Those who were recorded in the history of the previous three months were labeled as recent. On the other hand, those who came before this point in time were relegated to the past.

Table~\ref{tab:results} presents the empirical evidence that we gathered during our extensive experimentation. The highlights of our model's performance are arranged in columns and rows. By applying a critical eye to statistical analysis, we compared the many combinations of $\alpha$ and $\beta$ and uncovered the complex dynamics between these variables.

It is clear that tests considering the quantity of users who generated check-in records and the time passed since check-in data were generated outperformed the baseline in a range of scenarios. The scenario would have been different if the model had been trained for more epochs, even though baseline has the best \textbf{Acc@1}.

\begin{table*}[h]
    \centering
    \caption{Experimental Results}
    \begin{tabular}{|c|c|*{5}{c|}}
        % Header
        \hline
        \multicolumn{2}{|c|}{} & \textbf{Acc@1} & \textbf{Acc@5} & \textbf{Acc@10} & \textbf{Acc@20} & \textbf{MRR} \\

        % Baseline
        % Our Model
        \hline
        \multicolumn{7}{|c|}{\textbf{Baseline}} \\
        
        \hline
        \multicolumn{2}{|c|}{\textbf{GETNext}} & \textbf{0.6177} & 0.8477 & 0.9051 & 0.9455 & \textbf{0.7196} \\

        \hline
        \multicolumn{7}{|c|}{} \\

        % Our Model
        \hline
        \bm{$\alpha$} & \bm{$\beta$} & \multicolumn{5}{|c|}{\textbf{Our Proposed Methodology}} \\

        % alpha = 0.33
        \hline
        \multirow{3}{*}{\centering\textbf{0.33}} & \textbf{0.33} & 0.5856 & 0.8468 & 0.8946 & 0.9468 & 0.6984 \\
        \cline{2-7}
         & \textbf{0.50} & 0.5949 & \textbf{0.8622} & 0.9108 & 0.9470 & 0.7115 \\
        \cline{2-7}
         & \textbf{0.67} & 0.5995 & 0.8360 & 0.8889 & 0.9294 & 0.7030 \\
        \hline
        
        % alpha = 0.50
        \hline
        \multirow{3}{*}{\centering\textbf{0.50}} & \textbf{0.33} & 0.5268 & 0.7951 & 0.8586 & 0.9174 & 0.6481 \\
        \cline{2-7}
         & \textbf{0.50} & 0.5615 & 0.8174 & 0.8799 & 0.9323 & 0.6769 \\
        \cline{2-7}
         & \textbf{0.67} & 0.5721 & 0.8233 & 0.8792 & 0.9227 & 0.6820 \\
        \hline

         % alpha = 0.67
        \hline
        \multirow{3}{*}{\centering\textbf{0.67}} & \textbf{0.33} & 0.5346 & 0.7939 & 0.8534 & 0.9183 & 0.6485 \\
        \cline{2-7}
         & \textbf{0.50} & 0.5445 & 0.8228 & 0.8858 & 0.9328 & 0.6679 \\
        \cline{2-7}
         & \textbf{0.67} & 0.5988 & 0.8527 & \textbf{0.9110} & \textbf{0.9500} & 0.7147 \\
        \hline
        
    \end{tabular}
    \label{tab:results}
\end{table*}

\begin{table*}[h]
    \centering
    \begin{minipage}{\textwidth}
        \footnotesize\textit{Note: The GETNext model's output is shown first as baseline. The performance metrics for various combinations of $\alpha$ and $\beta$ parameters are shown in the subsequent rows. The exact values of $\alpha$ and $\beta$ for each matching row are shown in the first two columns, respectively.}
    \end{minipage}
\end{table*}

\section{Conclusion}
Our aim was to further the field of POI recommendation systems by redefining the definition of popularity ascribed to a point-of-interest (POI) within the limitations of this study. Our interpretation of POI popularity guarantees that recommended POIs represent the tastes of a larger, more diversified user base, rather than being biased by a small number of very active users, by taking into account both the check-in count and the number of unique users. Also, recommendations are kept up to date with the latest user behavior thanks to the integration of recent trends.

POI recommendation systems will be significantly impacted by this strategy. By avoiding biases brought about by excessively frequent check-ins from a small number of people, it improves fairness. To further represent the interests of the larger community, the model encourages recommendations that are more socially inclusive. We implement this new idea into a graph-based learning model called GETNext, which enables the scalable implementation of POI suggestions on a global graph structure. Strong and contextually relevant recommendations are ensured by the interplay between parameters \(\alpha\) and \(\beta\), as described in equation \ref{eq:pop}. This further refines the balance between user diversity and current trends.

Our investigation doesn't finish here; rather, it's just the beginning of a longer journey to learn more about the time aspect of POI recommendation. We will next attempt to understand the complex time dynamics at work, determine the time influence on other variables, and clarify how the more general travel patterns change with the seasons. A greater comprehension of the fundamental processes guiding POI recommendation systems is anticipated to be revealed by these upcoming studies, opening the door to more reliable and contextually sensitive recommendation frameworks.
%
% ---- Bibliography ----
%
\bibliographystyle{splncs04}
\bibliography{bibliography}

\end{document}